\documentclass[%
 aip,
 amsmath,
 amssymb,
 reprint,%
]{revtex4-1}

\newcommand{\ket}[1]{  \bigl | #1 \bigr \rangle}

\usepackage{amsfonts}
\usepackage{amssymb}
\usepackage{graphicx}
\usepackage{amsmath}
\usepackage{bm}
\DeclareMathAlphabet{\mathcal}{OMS}{cmsy}{m}{n}
\usepackage[english]{babel}
\usepackage{color}
\usepackage[version=3]{mhchem} 

\definecolor{darkgreen}{RGB}{40,110,5}

\usepackage{ulem}  %unter- (\uline{important}) und durchstreichen (\sout{wrong})
\usepackage{dcolumn}% Align table columns on decimal point
\usepackage{amsfonts}
\usepackage{amssymb}
\usepackage{graphicx}
\usepackage{amsmath}
\usepackage{bm}
\DeclareMathAlphabet{\mathcal}{OMS}{cmsy}{m}{n}
\usepackage[english]{babel}
\usepackage{color}
\usepackage[version=3]{mhchem} % Formula subscripts using \ce{}
\definecolor{darkgreen}{RGB}{40,110,5}

\usepackage{ulem}  %unter- (\uline{important}) und durchstreichen (\sout{wrong})
\definecolor{orange}{rgb}{1,0.5,0}

\newcommand{\ozg}[1]{{\color{orange}{}}}% \sout{#1}}}

\usepackage{dcolumn}% Align table columns on decimal point
\usepackage[utf8]{inputenc}
\usepackage[T1]{fontenc}
\usepackage{mathptmx}
\usepackage{etoolbox}

\makeatletter
\def\@email#1#2{%
 \endgroup
 \patchcmd{\titleblock@produce}
  {\frontmatter@RRAPformat}
  {\frontmatter@RRAPformat{\produce@RRAP{*#1\href{mailto:#2}{#2}}}\frontmatter@RRAPformat}
  {}{}
}%
\makeatother
\begin{document}

\title{Finite temperature dynamics in a polarized sub-Ohmic heat bath: a hierarchical equations of motion-tensor train study}
\author{Hideaki Takahashi}
\author{Raffaele Borrelli}
\affiliation{DISAFA, University of Torino,
Grugliasco, Italy}
\email{raffaele.borrelli@unito.it}
\author{Maxim F. Gelin}
\affiliation{School of Science, Hangzhou Dianzi University, Hangzhou 310018, China}
\email{maxim@hdu.edu.cn}
\author{Lipeng Chen}
\affiliation{Zhejiang Laboratory, Hangzhou 311100, China}
\email{chenlp@zhejianglab.com}

\begin{abstract}
Dynamics of the sub-Ohmic spin-boson model under polarized initial conditions at finite temperature is investigated by employing both analytical tools and the numerically accurate hierarchical equations of motion-tensor train method. By analyzing the features of nonequilibrium dynamics, we discovered a  bifurcation phenomenon which separates two regimes of the dynamics. It is found that before the bifurcation time, increasing temperature slows down the population dynamics, while the opposite effect occurs after the bifurcation time. The dynamics is highly sensitive to both initial preparation of the bath and thermal effects.

\end{abstract}

\maketitle

\section{Introduction}
Understanding microscopic mechanisms responsible for relaxation and decoherence of quantum systems coupled to their environments is the central task of open quantum systems. \cite{Leggett,Weiss} Due to the intricate interplay between electronic and vibrational degrees of freedom (DOFs), accurate simulation of the dynamics of dissipative systems is a formidable challenge. A common way for the evaluation of the dynamics of open quantum systems is based on quantum
master equations (QMEs). The key ingredients of this methodology are  tracing
out environmental DOFs and obtaining equations of motion for the reduced (system) density matrix.  \cite{BreuerBook} In many scenarios, however, the system-environment coupling is not weak, and dynamic timescales of the system and the environment  are comparable. As a result, traditional QMEs based on the Markovian  and/or weak system-bath coupling approximations  are no longer valid. \cite{LPReview,ChemRevLP} To capture the intrinsic non-perturbative and non-Markovian effects, several advanced numerically accurate methods and simulation protocols  have been developed. The quasiadiabatic propagator path integral (QUAPI), \cite{QUAPI1,QUAPI2} the hierarchy equations of motion (HEOM) \cite{HEOMTaka1,HEOMTaka2} and the path integral Monte Carlo (PIMC) \cite{PIMC1,PIMC2} are typical examples of  numerically exact QME approaches to the  propagation of the reduced density matrix.

Alternative methodology is based on the discretization of the bath DOFs and the numerical evaluation of the combined system-bath time dependent Schr\"odinger equation. Several formally exact wave function methods are available for the simulation of the dynamics at zero temperature. They include the multiconfiguration time-dependent Hartree (MCTDH) method \cite{MCTDHMethod} and its multilayer variant (ML-MCTDH), \cite{MLMCTDHMethod1,MLMCTDHMethod2,MLMCTDHMethod3} the time-dependent density matrix renormalization group (TD-DMRG), \cite{TDDMRGMethod1,TDDMRGMethod2,ChainMP1} the time-evolving block-decimation (TEBD) method, \cite{TEBD1,TEBD2,TEBD3}  the multiple Davydov Ansatz (mDA), \cite{zhao1,zhao2} the multi-configuration Ehrenfest,  \cite{MCEMethod1,MCEMethod2,MCEMethod3} and various variants of the matrix-product-state (MPS) methods. \cite{Verstraete13,Legeza15,Chin16,Reichman18,Schollwoeck19}
A promising technique to simulate finite-temperature dynamics with wave-function methods is based on  the so-called thermo-field dynamics-tensor train (TFD-TT) approach. \cite{TFDTTDyn2}

The spin-boson model (SBM), which describes two-level system coupled to the bath of harmonic oscillators, has established itself as perhaps one of the most widely studied models which facilitated understanding of relaxation and decoherence processes in open quantum systems. Despite its simple formulation, the SBM has attracted widespread attention over past years due to the rich physics it exhibits. It is thus not surprising that the dynamics of the SBM was simulated with virtually all aforementioned  methods. \cite{SBMQC1,SBMQC2,SBMSQ1,SBMSQ2,SBMND1,SBMND2,SBMET1,SBMET2,Bulla2003,Bulla2009,Fehske} Variational quantum algorithms were also employed to calculate the dynamics of the SBM. \cite{Tavernelli21}  

In the context of the present study, we wish to highlight several key results obtained on the dynamics of the SBM.  For the Ohmic bath with spectral density $J(\omega)\sim\omega^s$ and spectral exponent $s=1$,  increasing system-bath coupling turns the system dynamics incoherent and causes delocalized-localized phase transition at zero temperature. \cite{Leggett,Weiss} The sub-Ohmic bath with spectral exponent $s<1$ is characterized by the pronounced effect of low-frequency modes, which leads to strongly non-Markovian dynamics even for relatively weak system-bath interactions. Numerical simulations based on the ML-MCTDH method have shown that the dynamics of the SBM changes from the weakly damped coherent motion to localization when the system-bath coupling increases. \cite{WangH1,WangH2} Both QUAPI and MPS simulations have revealed that the initial preparation of the bath has considerable influence on the dynamics of the sub-Ohmic SBM. \cite{PNalbach, AlexChinMPSPRB2016} Using the PIMC method, it was further found that, under the polarized bath initial condition, the oscillatory dynamics persists irrespective of the system-bath coupling strength for $0<s<1/2$. \cite{KastPRL,KastPRB} For the factorized bath initial condition, the zero-temperature dynamic phase diagram containing coherent, pseudocoherent and incoherent phases was obtained by using the numerical technique based on the time-evolving matrix product operator.  \cite{OtterpohlPRL} In a recent work, the  zero-temperature dynamic phase diagrams of the sub-Ohmic SBM were identified  under various bath initial conditions by employing the mDA. \cite{LPSBMDiffInit}  The TFD-TT methodology was harnessed for studying finite-temperature effects on the dynamics of the SBM. \cite{TFDTTDyn1} 

In this paper, we delve into the still unexplored area of the dynamics of the sub-Ohmic SBM. By employing the HEOM in the twin-space (TS) representation and TT format, we perform dynamic simulations  of the sub-Ohmic SBM under the polarized bath initial condition.
We demonstrate that the HEOM-TSTT method is capable of accurately evaluating long-time dynamics of the sub-Ohmic SBM at any temperature. The rest of the paper is organized as follows. In Sec.~II, we introduce the SBM and the HEOM-TSTT method. In Sec.~III, we discuss temperature effects on the dynamics of the sub-Ohmic SBM. Conclusions are drawn in Sec.~IV. 

Throughout this paper, we set the reduced Planck constant $\hbar=1$ and the Boltzmann constant $k_B=1$, as well as employ  dimensionless variables. 

\section{Theory}

\subsection{Hamiltonian}

The Hamiltonian of the SBM can be written as 
\begin{equation}\label{H}
\hat{H}=\hat{H}_{\mathrm{S}}+\hat{H}_{\mathrm{B}}+\hat{H}_{\mathrm{SB}}
\end{equation}
where 
\begin{eqnarray}
\hat{H}_{\mathrm{S}}&=&-\frac{1}{2}\Delta\hat{\sigma}_x+\frac{1}{2}\epsilon\hat{\sigma}_z \nonumber \\
\hat{H}_{\mathrm{B}}&=&\sum_k\omega_k\hat{b}_k^{\dagger}\hat{b}_k \\
\hat{H}_{\mathrm{SB}}&=&\frac{1}{2}\hat{\sigma}_z\sum_k\lambda_k(\hat{b}_k^{\dagger}+\hat{b}_k)\nonumber 
\end{eqnarray}
For the spin system, $\hat{\sigma}_x$ and $\hat{\sigma}_z$ are Pauli matrices, and $\epsilon$ and $\Delta$ are the spin bias and the bare tunneling constant. For the bath, $\hat{b}_k^{\dagger}$ ($\hat{b}_k$) is the creation (annihilation) operator of the $k$th bosonic mode with frequency $\omega_k$. The system-bath interaction follows a bilinear form with $\lambda_k$ being the coupling strength. The bath is characterized by the spectral density function
\begin{equation}\label{Jw}
J(\omega)=\sum_k|\lambda_k|^2\delta(\omega-\omega_k)=2\alpha\omega_c^{1-s}\omega^se^{-\omega/\omega_c}.
\end{equation}
Here, $\alpha$ is the Kondo parameter specifying the  dimensionless coupling strength, $\omega_c$ is the cutoff frequency, and $s$ is the spectral exponent. The bath is categorized as sub-Ohmic ($0<s<1$), Ohmic ($s=1$) and super-Ohmic ($s>1$). In this paper, we limit our investigation to the sub-Ohmic bath which is characterized by strong non-Markovian effects induced by low frequency bath modes.  

\subsection{Shifted initial condition}\label{SBID}

Dynamics of the sub-Ohmic SBM is sensitive to the initial preparation of the bath. \cite{PNalbach,KastPRL,WangLu} 
Two kinds of bath initial conditions are usually considered, the so-called factorized initial condition and the shifted (polarized) initial condition. For the factorized initial condition, the spin system is in thermal equilibrium with the bath at temperature $T$ and at the initial time $t=0$, i.e., 
\begin{equation}\label{normal}
\hat{\rho}(t=0)=\hat{\rho}_{\mathrm{S}}(t=0)\otimes\hat{\rho}_{\mathrm{B}}
\end{equation}
Here, $\hat{\rho}(t=0)$ and $\hat{\rho}_{\mathrm{S}}(t=0)$ are initial total and system density matrices, respectively, and 
\begin{equation}\label{Bol}
\hat{\rho}_{\mathrm{B}}=e^{-\beta\hat{H}_{\mathrm{B}}}/\mathrm{Tr}\{{e^{-\beta\hat{H}_{\mathrm{B}}}}\}
\end{equation}
is the equilibrium (Boltzmann) bath distribution at the inverse temperature $\beta=1/T$. 

For the shifted (polarized) initial condition, one prepares the spin long time before the bath can thermalize to the shifted spin, i.e.,  
\begin{equation}\label{shifted}
\hat{\rho}(t=0)=\hat{\rho}_{\mathrm{S}}(t=0)\otimes\hat{\rho}_{\mathrm{B\mu}}
\end{equation}
with 
\begin{equation}
\hat{\rho}_{\mathrm{B\mu}}=e^{-\beta(\hat{H}_{\mathrm{B}}+\hat{H}_{\mathrm{SB}}|_{\hat{\sigma}_z=\mu})}/\mathrm{Tr}\{e^{-\beta(\hat{H}_{\mathrm{B}}+\hat{H}_{\mathrm{SB}}|_{\hat{\sigma}_z=\mu})}\}
\end{equation}
where $\mu$ is a parameter characterizing the degree of the bath polarization. In particular, $\mu=0$ corresponds to the factorized initial condition of Eq.~(\ref{normal}). In the present work, we explore the impact of  the shifted initial condition of Eq.~(\ref{shifted}) on the dynamics of the sub-Ohmic SBM at finite temperature.

\subsection{Transformed Hamiltonian}\label{TranH}

The dynamics of the SBM under factorized and shifted initial conditions are closely related. To demonstrate that, we formulate the following statement,

\textit{Theorem}. The Liouville-von Neumann (LvN) equation   \begin{equation}
\label{eq:liouville}
\partial_t\hat{\rho}(t)=-i[\hat{H},\hat{\rho}(t)]
\end{equation}
governed by the SBM Hamiltonian $\hat{H}$ under the  shifted initial condition of Eq. (\ref{shifted}) is equivalent to the LvN equation 
\begin{equation}\label{mrho}
\partial_t\hat{\rho}(t)=-i[\hat{H}_{\mu},\hat{\rho}(t)]\end{equation}
governed by the modified SBM Hamiltonian $\hat{H}_{\mu}$ under the factorized initial condition of Eq. (\ref{normal}). 
 Here the modified SBM  Hamiltonian 
\begin{equation}\label{modH}
\hat{H}_{\mu}=\hat{H}_{\mathrm{S}}^{\mu}+\hat{H}_{\mathrm{B}}+\hat{H}_{\mathrm{SB}}^{\mu}
\end{equation}
has the same structure as the original SBM Hamiltonian $\hat{H}$, but its parameters become $\mu$-dependent:
\begin{eqnarray}\label{Hmod}
\hat{H}_{\mathrm{S}}^{\mu}=-\frac{\Delta}{2}\hat{\sigma}_x+\frac{\epsilon-\mu{E}_{\mathrm{R}}}{2}\hat{\sigma}_z, \\
\hat{H}_{\mathrm{SB}}^{\mu}=\frac{\hat{\sigma}_z-\mu}{2}\sum_k\lambda_k(\hat{b}_k^{\dagger}+\hat{b}_k).
\end{eqnarray}
Here 
\begin{equation}\label{ER}
E_R=\int_0^{\infty}\frac{J(\omega)}{\omega}d\omega=2\alpha\omega_c\Gamma(s)  
\end{equation}
is the reorganization energy and $\Gamma(s)$ denotes the  Gamma function. 

\textit{Proof}. We can rewrite the Hamiltonian 
$$
\hat{H}_{\mathrm{B}\mu}=\hat{H}_{\mathrm{B}}+\hat{H}_{\mathrm{SB}}|_{\hat{\sigma}_z=\mu}
$$
as 
\begin{eqnarray}
\hat{H}_{\mathrm{B}\mu}=&&\sum_k\omega_k\hat{b}_k^{\dagger}\hat{b}_k+\frac{\mu}{2}\sum_k\lambda_k(b_k^{\dagger}+\hat{b}_k)\nonumber \\
=&&\sum_k\omega_k(\hat{b}_k^{\dagger}+\mu\frac{\lambda_k}{2\omega_k})(\hat{b}_k+\mu\frac{\lambda_k}{2\omega_k})-\mu^2\sum_k\frac{\lambda_k^2}{4\omega_k}\nonumber \\
=&&\hat{H}_{\mathrm{B}\mu}^{\mathrm{R}}-\frac{\mu^2}{4}E_{\mathrm{R}}
\end{eqnarray}
where 
$$
\hat{H}_{\mathrm{B}\mu}^{\mathrm{R}}=\sum_k\omega_k(\hat{b}_k^{\dagger}+\mu\frac{\lambda_k}{2\omega_k})(\hat{b}_k+\mu\frac{\lambda_k}{2\omega_k}).
$$
Obviously, the Boltzmann distributions corresponding to $\hat{H}_{\mathrm{B}\mu}$ and $\hat{H}_{\mathrm{B}\mu}^{\mathrm{R}}$ are identical, 
\begin{equation}
\hat{\rho}_{\mathrm{B}\mu}=\frac{e^{-\beta\hat{H}_{\mathrm{B}\mu}}}{\mathrm{Tr}\{e^{-\beta\hat{H}_{\mathrm{B}\mu}}\}}=\hat{\rho}_{\mathrm{B}\mu}^{\mathrm{R}}=\frac{e^{-\beta\hat{H}^{\mathrm{R}}_{\mathrm{B}\mu}}}{\mathrm{Tr}\{e^{-\beta\hat{H}^{\mathrm{R}}_{\mathrm{B}\mu}}\}}.
\end{equation}
Let us now introduce the new shifted creation (annihilation) operator for the $k$th bosonic mode
\begin{equation}
\hat{c}_k^{\dagger}=\hat{b}_k^{\dagger}+\mu\frac{\lambda_k}{2\omega_k},\quad\hat{c}_k=\hat{b}_k+\mu\frac{\lambda_k}{2\omega_k}
\end{equation}
(the transformation is equivalent to the polaron transformation  \cite{Weiss} of the original Hamiltonian). In the representation of the new operators, the Boltzmann distribution $\hat{\rho}_{\mathrm{B}\mu}(\hat{b}_k^{\dagger},\hat{b}_k)$  corresponds to the factorized initial condition, i.e., $\hat{\rho}_{\mathrm{B}\mu}(\hat{b}_k^{\dagger},\hat{b}_k)\Leftrightarrow\hat{\rho}_{\mathrm{B}}(\hat{c}_k^{\dagger},\hat{c}_k)$. In the new representation, the original Hamiltonian $\hat{H}$ has the form 
\begin{eqnarray}
\label{eq:originalHam}
\hat{H}=&&-\frac{\Delta}{2}\hat{\sigma}_x+\frac{\epsilon}{2}\hat{\sigma}_z+\sum_k\omega_k\hat{b}_k^{\dagger}\hat{b}_k+\frac{\hat{\sigma}_z}{2}\sum_k\lambda_k(\hat{b}_k^{\dagger}+\hat{b}_k)\nonumber \\
=&&-\frac{\Delta}{2}\hat{\sigma}_x+\frac{\epsilon}{2}\hat{\sigma}_z+\sum_k\omega_k(\hat{c}_k^{\dagger}-\mu\frac{\lambda_k}{2\omega_k})(\hat{c}_k-\mu\frac{\lambda_k}{2\omega_k})\nonumber \\
&&+\frac{\hat{\sigma}_z}{2}\sum_k\lambda_k(\hat{c}_k^{\dagger}-\mu\frac{\lambda_k}{2\omega_k}+\hat{c}_k-\mu\frac{\lambda_k}{2\omega_k})\nonumber \\
&&=-\frac{\Delta}{2}\hat{\sigma}_x+\frac{\epsilon-\mu{E}_{\mathrm{R}}}{2}\hat{\sigma}_z+\sum_k\omega_k\hat{c}_k^{\dagger}\hat{c}_k\nonumber \\
&&+\frac{\hat{\sigma}_z-\mu}{2}\sum_k\lambda_k(\hat{c}_k^{\dagger}+\hat{c}_k)+\frac{\mu^2}{4}E_{\mathrm{R}}
\end{eqnarray}
The above expression yields the Hamiltonian of Eq.~(\ref{modH}), in which we dropped the constant term of $\mu^2E_{\mathrm{R}}/4$ since it does not affect any dynamics.

Mapping of the shifted SBM to the factorized one facilitates numerical simulations. It demonstrates that the dynamics of the SBM under shifted initial conditions can be obtained with any method developed for the standard factorized initial conditions.  One just has to use the modified Hamiltonian $\hat{H}_{\mu}$ instead of $\hat{H}$. This is employed in the present work for the construction of the HEOM-TSTT integrator for the SBM under shifted initial conditions.

\subsection{HEOM in twin space}

In the following we will make use of the TS  formulation of the LvN equation. The reader is referred to the original papers for the comprehensive discussion on the topic, \cite{Borrelli2019JCP,BorrelliGelin2017SR,UmezawaEtAl1982,Suzuki1996IJMPB} but for the subsequent presentation it is enough to realize that the LvN equation (\ref{mrho})  in the TS representation takes the form 
\begin{equation}
\label{Twin}
\partial_t\ket{\rho(t)}=-i(\hat{H}_{\mu}-
\tilde{\hat{H}}_{\mu})\ket{\rho(t)}.
\end{equation}
Here $\tilde{\hat{H}}_{\mu}$ is a copy of the Hamiltonian $\hat{H}_{\mu}$ acting on the so-called {\it tilde} Hilbert  space, which is an exact copy of the original Hilbert space where the system dynamical variables are defined.  
Eq. (\ref{Twin}) is the departing point for the derivation of the HEOM-TSTT, as recently proposed by Borrelli. \cite{Borrelli2019JCP} 
Below we briefly sketch the main steps of the derivation, and further details can be found in Refs.  \cite{Borrelli2019JCP,BorrelliDolgov2021JPCB,ShiEtAl2018JCP}

First of all we represent Eq. (\ref{Twin}) in the interaction representation. Then we notice that the time evolution of the reduced density matrix governed by the modified Hamiltonian of Eq. (\ref{modH}) can be evaluated  as
\cite{Kubo1962JPSJ,IshizakiEtAl2010PCCP}
\begin{equation}
%\label{eq:tdsk}
	\ket{\rho_S(t)}_I = T_+ \exp\left(-\int_0^t \hat K_I^{(2)}(s)ds\right)\ket{\rho(0)}_I
\end{equation}
where
\begin{equation}
	\hat K_I^{(2)}(s)=
	\int_{0}^{s} d\tau \langle \hat H_{\mathrm{SB}}^{\mu}(s)
 \hat H_{\mathrm{SB}}^{\mu}(\tau)\rangle, 
\end{equation}
$T_+$ is the time-ordering operator, and the subscript $I$ indicates the interaction representation.
Starting from the modified Hamiltonian of Eq. (\ref{modH}), after some easy manipulations, it is possible to write the second order cumulant in the form
\begin{align}
\label{eq:k2corr}
	\hat K_I^{(2)}(t) = & \sum_k [S(t)-\tilde S(t)][C_{R}(t-s) (S(s)-\tilde S(s)) \nonumber \\ 
 & +  C_{I}(t-s) (S(s)+\tilde S(s))]
	%\hat K_I^{(2)}(t) = \sum_k [S_k(t)-\tilde S_k(t)][C_k(t-s) S_k(s)	- C_k(t-s)^* \tilde S_k(s)]
\end{align}
where $C_{R}(t)=(C(t)+C^*(t))/2$, $C_{I}(t)=(C(t)-C^*(t))/2$ are the symmetric and antisymmetric part of the bath correlation function (BCF) $C(t)$, respectively, 
$S = (\sigma_z-\mu)/2$ is the system operator, and $\tilde S$ is the \textit{tilde} counterparts.

Following the standard theory of HEOM we assume that both BCF components 
can be decomposed into a sum of exponential functions 
\begin{equation}
\label{eq:bcfexp}
	C_X(t-s) = \sum_{j=1}^{J_X} c_{Xj} e^{-\gamma_{Xj} |t-s|}
\end{equation} 
where $c_{Xj}$ and $\gamma_{Xj}$ with $X=R,I$ are complex numbers, and $J_X$ is the number of terms for the expansion of the $X$ component of $C(t)$.
This decomposition is crucial for disentangling the time-ordering in the time-evolution of the reduced density matrix. The set of coefficients and exponents can be obtained using a variety of techniques whose choice depends on the type of the spectral density.\cite{HuEtAl2011JCP,HuEtAl2010JCP,XuEtAl2022PRL}

The key to the derivation of the HEOM-TSTT  is the definition of  the so-called auxiliary density vectors (ADV)\cite{TanimuraKubo1989JPSJ}
\begin{gather}
	\ket{\rho_S^{\mathbf{mn}}(t)}=T_+ 
	 \prod_{j} (m_{j}!c_{Rj}^{m_{j}})^{-1/2} \nonumber \\
  \left(i\int_0^t ds
e^{-\gamma_{Rj}|t-s|} (S_{Rj}(s)- \tilde S_{Rj}(s))\right)^{m_{j}} \nonumber \\
\times \prod_{j} (n_{j}!(c_{Ij})^{n_{j}})^{-1/2}\left(i\int_0^t ds
e^{-\gamma_{Ij}^*|t-s|}(S_{Ij}(s)+\tilde S_{Ij}(s))\right)^{n_{j}}\nonumber \\
\times \exp\left(-\int_0^t \hat K_I^{(2)}(s)ds\right)\ket{\rho_S(0)}
\end{gather}
where $\ket{\rho_S(0)}$ is the reduced density matrix of the system, $\mathbf{m}=\{m_{j}\}$, $\mathbf{n}=\{n_{j}\}$
are two sets of non-negative integers and  
\begin{align}
    S_{Xj} =  \sqrt{c_{Xj}} S, \qquad X=R,I.
\end{align}
In the above formalism we have $J_R+J_I$ auxiliary spectral 
density vectors.
It is readily verified that the vector $\ket{\rho_S(t)}_I$, describing the physical state of our system, corresponds to the auxiliary state vector having all indices $(m_{j},n_{j})=0$, \textit{i.e.} $\ket{\rho_S(t)}_I=\ket{\rho_S^{\boldsymbol{0}}(t)}$. The above definition takes into account the scaling originally proposed by Shi and coworkers which improves the numerical stability of the final system of equations. \cite{ShiEtAl2009JCP}

The HEOM are readily derived upon repeated differentiation of $\ket{\rho_S^\mathbf{m n}}$ with respect to time and then moving to the Schr\"odinger representation. The final result can be written in the form
\begin{align}
\label{eq:heom}
	\frac{\partial}{\partial t}\ket{ \rho_S^\mathbf{mn}} =&  -\big(i\hat H_S^{\mu}	+\sum_{j} m_{j}\gamma_{Rj}+\sum_{j}n_{j}\gamma_{Ij}
 \big)\ket{\rho_S^\mathbf{mn}}  \nonumber \\ 
	&-i\sum_{j} \sqrt{m_{j}+1}(S_{Rj} -\tilde S_{Rj})
	(\ket{\rho_S^{\mathbf{m}+1_{j}~\mathbf{n}}} \nonumber \\
	&-i\sum_{j} \sqrt{m_{j}}(S_{Rj} -\tilde S_{Rj})
	\ket{\rho_S^{\mathbf{m}-1_{j}~\mathbf{n}}} 
 \nonumber \\
	&-i\sum_{j} \sqrt{n_{j}+1}(S_{Ij}-\tilde S_{Ij}) 
	\ket{\rho_S^{\mathbf{m}~\mathbf{n}+1_{j}}}
 \nonumber \\
	&+\sum_{j} \sqrt{n_{j}}(S_{Ij}+\tilde S_{Ij}) 
	\ket{\rho_S^{\mathbf{m}~\mathbf{n}-1_{j}}}.
\end{align}
The HEOM consist of an infinite set of first-order ordinary differential equations. 
Fortunately, it is possible to devise very efficient truncation schemes which allow us to obtain highly accurate results for the reduced system dynamics. 
The reader is referred to the original papers for the description of the optimal truncation scheme. \cite{TanimuraKubo1989JPSJa,IshizakiTanimura2005JPSJ} In the above derivation we have not considered low-temperature corrections which can be included straightforwardly from a direct application of the original approach suggested by Ishizaki and Tanimura.\cite{IshizakiTanimura2005JPSJ}

The HEOM-TSTT equations resemble their HEOM counterparts  in the standard density matrix  formalism (see, for example, Ref. \cite{IshizakiEtAl2010PCCP}). However, in the present case, commutators and anti-commutators are replaced by differences and sums of the physical and tilde   operators. As we shall see in the next section, this has several benefits for the numerical implementation of the HEOM solver. 

To further simplify the HEOM-TSTT structure, we introduce two sets of boson-like creation-annihilation operators 
$b_{Rj}^+,b_{Ij}^-$ and 
$\bar b_{Rj}^+,\bar b_{Ij}^-$ acting respectively on the set of state vectors 
$\ket{\mathbf{m}} = \ket{m_{0}...m_{J_R}}$, and
$\ket{\mathbf{n}} = \ket{n_{0}...n_{J_I}}$, 
\begin{equation}
	b_{Rj}^+\ket{\mathbf{m}} = \sqrt{(m_{j}+1)} \ket{\mathbf{m}+1_{j}}, \quad
	b_{Rj}^-\ket{\mathbf{m}} = \sqrt{m_{j}}\ket{\mathbf{m}-1_{j}} \quad
\end{equation}
\begin{equation}
\label{eq:aux}	\bar{b}_{Ij}^+\ket{\mathbf{n}} = \sqrt{(n_{j}+1)} \ket{\mathbf{n}+1_{j}}, \quad
	\bar{b}_{Ij}^-\ket{\mathbf{n}} = \sqrt{n_{j}}\ket{\mathbf{n}-1_{j}} 
\end{equation}
we then define the vector 
\begin{equation}
\label{eq:vecpsi}
\ket{\rho(t)} = \sum_\mathbf{mn} \ket{\rho_S^{\mathbf{mn}}(t)}\ket{\mathbf{mn}}	
\end{equation}
and rewrite the HEOM-TSTT  in the compact form
\begin{align}
\label{eq:heom2}
	\frac{\partial}{\partial t}\ket{\rho} =& 
 \bigg[-i\hat H_S^{\mu}	
 -\sum_{j,X=R,I} \gamma_{Xj} b^+_{Xj}b_{Xj}^-
 \nonumber \\ 
 &-i\sum_{j,X=R,I} (S_{Xj} -\tilde S_{Xj}) b_{Xj}^-		
	\nonumber \\
&-i\sum_{j}(S_{Rj}-\tilde S_{Rj})  b_{Rj}^+
+\sum_{j}(S_{Ij}+\tilde S_{Ij}) \bar{b}_{kj}^+
	\bigg]\ket{\rho},
\end{align}
with the initial condition given by $\ket{\rho(0)} = \ket{\rho_S(0)}\ket{\boldsymbol{0}}$.
%In the following the super-operator on the right-hand side of Eq. \ref{eq:heom2} will be referred to as $L_{HEOM}$.
%%%%%%%%%%%%% END HEOM

Finally, we mention that the TS formulation of HEOM has been recently extended to treat the interaction of molecules with fermionic baths at finite temperature.
The approach is very similar to the one introduced  above, and the reader is referred to the original papers for mathematical details. \cite{KeEtAl2022JCP,KeEtAl2023JCP}

%%%%%%%%%%%%%%%%%%%%%%%%%%%%%%% 

\subsection{TT representation of auxiliary density vectors\label{TSTTHEOM}}

The size of the set of equations (\ref{eq:heom2}) 
increases exponentially with the number of expansion terms of the spectral density and, besides, the equations
are known to be rather unstable. \cite{DunnEtAl2019JCP} The numerical solution is usually tackled by 
truncating the hierarchy in such a way that all ADVs with $\sum_k n_{ik} > D$ are 
set to zero, where $D$ is a prescribed integer. Typical values of $D$ range from 4 to 10, depending on the degree of non-Markovianity of the bath.
While this truncation scheme is rather effective and can significantly reduce the computational burden of the numerical solution of HEOM, \cite{KramerEtAl2018JCC}
it cannot solve the inherent exponential scaling of the number of equations. 
Only very recently, Shi \cite{ShiEtAl2018JCP} and later Borrelli \cite{Borrelli2019JCP} have shown that it is possible to use TT 
techniques to overcome this problem and significantly expand the range of applications of the HEOM methodology. 
%Meanwhile other types of tensor network approaches have also been suggested\cite{YanEtAl2020JCP,YanEtAl2021JCP,Ke2023JCP}. 

In the following we sketch the basic principles of the TT decomposition, 
\cite{LubichEtAl2015SJNA,LubichOseledets2013BNM,OseledetsTyrtyshnikov2009SJSC,Oseledets2011SJSC,HoltzEtAl2011NM,ThomasCarrington2015JPCA,DolgovEtAl2012SJSC,DolgovSavostyanov2014SJSC} and show how it can be applied to solve the HEOM-TSTT equation (\ref{eq:heom2}).

\begin{figure}
\begin{center}
% the figure has been obtained with the mptikz library.
%The pdf file is then converted using pstopdf -eps tt1.pdf figure1.eps
\includegraphics[width=8cm]{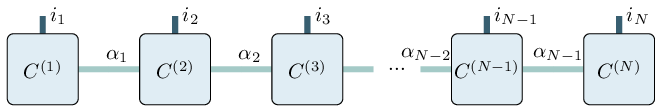}
\caption{\label{fig:tensortrain}
Graphical representation of a TT. Each square node represents a core of the TT, and each vertical line represents an index $i_k$ of the tensor. Connecting lines represent the contractions over the indices $\alpha_k$. }
\end{center}
\end{figure}

Let us first recall the basic principles of the TT decomposition by considering a generic expression of a state of a $N$-dimensional quantum system at time $t$ in the form
\begin{equation}
\label{eq:rhosumtt}
	\ket{\rho(t)} = \sum_{i_1,i_2,...,i_N} C_t(i_1,...,i_N)\ket{i_1}\otimes\ket{i_2}\cdots\ket{i_N}.
\end{equation}
where $\ket{i_k}$ labels the basis states of the $k$th dynamical variable, and the elements $C_t(i_1,...,i_N)$ are complex numbers labeled by $N$ indices. 
 If we truncate the summation of each index $i_k$ to a maximum value $p_k$, the elements $C_t(i_1,...,i_N)$ represent a tensor of order $N$. Evaluation of the summation in Eq. (\ref{eq:rhosumtt}) requires  computation (and storage) of $p^N$ terms with $p$ being the average size of the one-dimensional basis set, which becomes prohibitive for large $N$. Using the TT format, the tensor $C_t$ is approximated as
\begin{equation}
\label{eq:ttformat0}
C_t(i_1,...,i_N) \approx C^{(1)}(i_1)C^{(2)}(i_2)\cdots C^{(N)}(i_N)	
\end{equation}
where $C^{(k)}(i_k)$ is a $r_{k-1}\times r_k$ complex matrix, $k=1,\ldots,N$ (for the moment, let us drop the time variable for simplicity). In the explicit index notation
\begin{equation}
	\label{eq:ttformat}
	C_t(i_1,...,i_N) \approx \sum_{\alpha_0, \cdots,\alpha_N=1}^{r_0,\ldots,r_N} C^{(1)}_{\alpha_0,\alpha_1}(i_1)
	C^{(2)}_{\alpha_1,\alpha_2}(i_2)\cdots
	C^{(N)}_{\alpha_{N-1},\alpha_N}(i_N).
\end{equation}
The trailing indices $\alpha_0$ and $\alpha_N$ are introduced for uniformity of notation, but to render the right-hand side scalar, we always set $r_0=r_N=1$.
The factors $C^{(k)}$ are three-dimensional arrays, called 
\textit{cores} of the TT decomposition. The dimensions $r_k$ are called TT ranks. 
The TT decomposition~(\ref{eq:ttformat0}) is also known under the name of the MPS. \cite{FannesEtAl1992CP,Perez-GarciaEtAl2007QIC}
In the MPS language, the TT ranks are referred to as \textit{bond dimensions}.
Using the TT decomposition of Eq. (\ref{eq:ttformat0}) it is possible, at least in principle, to overcome most of the difficulties caused by the dimension of the problem. Indeed, the wave function is entirely defined by $N$ arrays of dimensions $r_{k-1}\times p_k \times r_{k}$ thus requiring a storage dimension of the order $Npr^2$.
%%%
The TT-decomposition is visualized in  Fig.~\ref{fig:tensortrain}.

Turning now to the representation of the ADV of Eq.~(\ref{eq:vecpsi}) in the TT format we let $d$ be the number of DOFs of the Hamiltonian operator  $\hat{H}_S^{\mu}$, and assume that the dissipative environment is described using  $M$ uncorrelated spectral densities $J_k(\omega)$ each expanded into $J$ Matstubara terms. 
Hence, the  ADV  can be considered as a tensor with $N=2(d+JM)$ indices. Therefore, in Eq. (\ref{eq:ttformat}) the first $2d$ indices  label the physical and tilde DOFs of the system,
and the remaining $2JM$ indices label the bath operators.

\subsection{Fitting of the sub-Ohmic spectral density}

The expansion of the symmetric ($X=R$) and antisymmetric ($X=I$)  part of the BCF $C_{X}(t)$ using equation (\ref{eq:bcfexp}) is a complex task that strongly affects the accuracy of the solution, for which many procedures have been developed.\cite{HEOMTaka2} 
In this work we have used the so-called ESPRIT fitting
methodology, \cite{RoyKailath1989ITASSP,PottsTasche2013LAaiA,AnderssonCarlsson2018SJMA&A} which is a direct time-domain fitting procedure. 
For a given BCF component $C_X(t)$ with $X=R,I$, we consider the vector $\mathbf{f}=\left(f_0, f_1, \ldots, f_{2N-2}\right)^T$ where each element $f_j=C_X(t_j)$ is sampled on an equispaced grid $t_j= hj,(h=\frac{t_\mathrm{max}}{2N-2},\, j=0,1, \ldots, 2N-2)$ and $N>M$.  Here we assume that $f_j$ can be expressed as
\begin{equation}
    f_j=\sum_{k=1}^M c_k e^{-a_khj}=\sum_{k=1}^M c_k z_k^j,
\end{equation}
where $c_k\in \mathbb{C}$ and $z_k=e^{-a_k h}\in \mathbb{D}$, and the symbol $\mathbb{D}$ denotes the complex unit disk without zero. Accordingly, the problem is reduced to a problem of finding complex weights $\boldsymbol{c}=\left(c_1, \ldots, c_M\right)^T$ and complex nodes $\boldsymbol{z}=\left(z_1, \ldots, z_M\right)^T$ with a minimal number of terms $M$ for a given accuracy $\epsilon$, which is written as
\begin{equation}
    \left\|f_j-\sum_{k=1}^M c_k z_k^j\right\| < \epsilon
\end{equation}
for all $j=0,\dots,2N-2$.  ESPRIT algorithm relies on the rank reduction of the Hankel matrix $H_{N,N}=(f_{k+l})_{k,l=0}^{N-1,N-1}$, which has $f_j$ as entries, and the overdetermined least-squares Vandermonde system.  While ESPRIT is similar to Prony's method\cite{BeylkinMonzon2005AaCHA,ChenEtAl2022JCP} in principle, it offers greater stability and improved accuracy.  In this study, we refer to Algorithm 1 in Ref.~\onlinecite{AnderssonCarlsson2018SJMA&A} and use the routine \textsf{DQAG} from the \textsf{QUADPACK} library with an accuracy of $\epsilon_\mathrm{DQAG}=10^{-12}$ to obtain highly accurate data of $\mathbf{f}$ through the quantum fluctuation-dissipation theorem.

\section{Results and discussion}\label{Res}

In this section, we systematically study the spin dynamics $P_z(t)$ for different values of the system-bath coupling $\alpha$ and the bath exponent $s$ at various temperatures $T$. We begin with analytical results, discussing the short-time expansion of $P_z(t)$ in Sec. \ref{short}. Then we present results of the HEOM-TSTT simulations  in Sec. \ref{sim}. 
In all HEOM-TSTT calculations, we set $\Delta=0.1$ and $\epsilon=0$  as well as consider the strongly polarized initial condition ($\mu=1$). The cutoff frequency of the bath is chosen as the unit of energy, $\omega_c=1$. %The dimension of the local phonon Hilbert space is fixed at 3 and up to 120 phonon modes are kept to reach the convergence of the results.

\subsection{Short-time expansion of $P_{z}(t)$}\label{short}

The simulation of the dynamics of the SBM at zero temperature  with the spectral density of Eq.~(\ref{Jw}) revealed the following rule-of-thumb principle.  
The time evolution of $P_{z}(t)$ calculated for a series of coupling strengths 
$\alpha$ or bath exponents $s$ at zero temperature do not intersect  each other up to quite long times $t$ (see, e.g., Refs.  \cite{WangH2,OtterpohlPRL,LPSBMDiffInit}). This observation hints that the inspection of the  short-time
behavior of $P_{z}(t)$  can be useful for the analyses of $P_{z}(t)$. 

The transformation of Sec.~\ref{TranH} can be  conveniently  used  for performing the Taylor-series
expansion of $P_{z}(t)$, because the evaluation of necessary expectation values with the equilibrium Boltzmann distribution of Eq.~(\ref{Bol}) can be done straightforwardly with standard methods. We thus need to evaluate 
\[
P_{z}(t)=\mathrm{Tr}\{\hat{\sigma}_z\hat{\rho}(t)\}=\sum_{a=0}^{\infty}\mathrm{Tr}\{\hat{\sigma}_z\hat{\rho}_a\}\frac{t^{a}}{a!},
\]
where 
\[
\hat{\rho}_0=|\uparrow\rangle\langle\uparrow|\hat{\rho}_{\mathrm{B}}, \,\,\, \hat{\rho}_{a+1}=-i[\hat{H}_{\mu},\hat{\rho}_{a}].
\]
A tedious but straightforward calculation yields the    following result,
\begin{equation}\label{Pz}
P_{z}(t)=1-2\Delta^{2}t^{2}+\Delta^{2}\left(\Omega_{\mu}^{2}+\varUpsilon\right)t^4/6+O(t^{6}).
\end{equation}
Here 
\begin{equation}\label{Rabi}
\Omega_{\mu}=\sqrt{(\varepsilon+\mu E_{\mathrm{R}})^{2}+4\Delta^{2}}
\end{equation}
 is the $\mu$-renormalized Rabi frequency, and 
\begin{equation}\label{Yp}
\varUpsilon=\int d\omega J(\omega)\coth\left(\frac{\omega}{2T}\right).
\end{equation}
According to Eq.~(\ref{ER}), the reorganization energy $E_{\mathrm{R}}$ increases with $\alpha$, decreases with $s$, and is temperature-independent. As a result, the Rabi-frequency $\Omega_{\mu}$ is also temperature-independent, but depends significantly on the system  preparation (parameter $\mu$), as well as increases with $\alpha$ and decreases with $s$. 
For the bath spectral density of Eq.~(\ref{Jw}),  $\varUpsilon$ cannot be evaluated in the closed form for finite $T$.
It is crucial though that it is strictly positive, increases with  temperature (it can immediately be verified that $d\varUpsilon/dT>0$) as well as grows  with the coupling strength $\alpha$. As a function of the spectral exponent $s$, $\varUpsilon$ is U-shaped, exhibiting a $T$-dependent minimum.

The expansion of Eq. (\ref{Pz}) contains only even powers of time owing to the time reversibility of the Hamiltonian dynamics.    
The term $\sim t^2$ is negative and is exclusively determined by the spin system, that is, by the tunneling $\Delta$. This term describes the short-time
dynamics and specifies the so-called Zeno time $\tau_{Z}=\Delta^{-1}/\sqrt{2}$, meaning that the initial depopulation can be approximated as  
$$
P_{z}(t)\approx \exp[-(t/t_Z)^2]+O(t^4)
$$
(see, e.g. Refs. \cite{Zeno01,Zeno19}).  
The term $\sim t^4$ is strictly positive and depends on the bath parameters and  temperature. On the basis of the scaling properties of $E_{\mathrm{R}}$ and  $\varUpsilon$ established above, we can therefore conclude that  $P_{z}(t)$
increases with $\alpha$ and $T$. 
In Sec. \ref{Res} B, we will explore how these short-time results can be used to help understanding global $P_{z}(t)$ dynamics at finite temperature and polarized initial conditions.

Note that $\varUpsilon$ can be analytically evaluated at zero temperature. The result reads as $\varUpsilon=2\alpha\omega_{c}^{2}\Gamma(s+1)$. It is known that $\Gamma(s)$ decreases
with $s$ for $0<s<1.462$, reaches its minimum at $\Gamma(1.462)=0.886$
and then increases for $s>1.462$. Taking into account that
$E_{\mathrm{R}}\sim\Gamma(s)$ according to Eq.~(\ref{ER}), $P_{z}(t)$ decreases with
$s$ for $0<s<0.462$ and increases for $s>1.462$. The behavior for
$0.462<s<1.462$ is determined by the interplay of $\Omega_{\mu}^{2}$
(decreases with $s$) and $\varUpsilon$ (increases with
$s$), which probably explains why the value of $s\approx0.46$ is the bifurcation point in the $s-\alpha$ phase diagrams. \cite{OtterpohlPRL,LPSBMDiffInit} 

Eq. (\ref{Pz}) can be used to anticipate the microscopically justified fitting functions for $P_{z}(t)$. One of the possible expression reads 
\begin{equation}\label{RTT}
P_{z}(t)=1-4\frac{\Delta^{2}}{\Omega_{\mu}^{2}(T)}\left(1-\cos(\Omega_{\mu}(T) t)\right)+O(t^6)
\end{equation}
where 
\begin{equation}\label{RT}
\Omega_{\mu}(T)=\sqrt{\Omega_{\mu}^{2}+\varUpsilon}    
\end{equation}
is the effective temperature-dependent Rabi frequency. 
This formula is inspired by the following observation, for bath-free SBM ($E_{\mathrm{R}}=\varUpsilon=0$) we recover the standard Rabi formula. Since  $\varUpsilon$ increases with  temperature, Eq.~(\ref{RT}) predicts that temperature increases the effective Rabi frequency $\Omega_{\mu}(T)$
and therefore shortens the period of Rabi oscillations.

\subsection{HEOM-TSTT simulations of $P_{z}(t)$}\label{sim}

\begin{figure}
    \centering
    \includegraphics[width=1.0\linewidth]{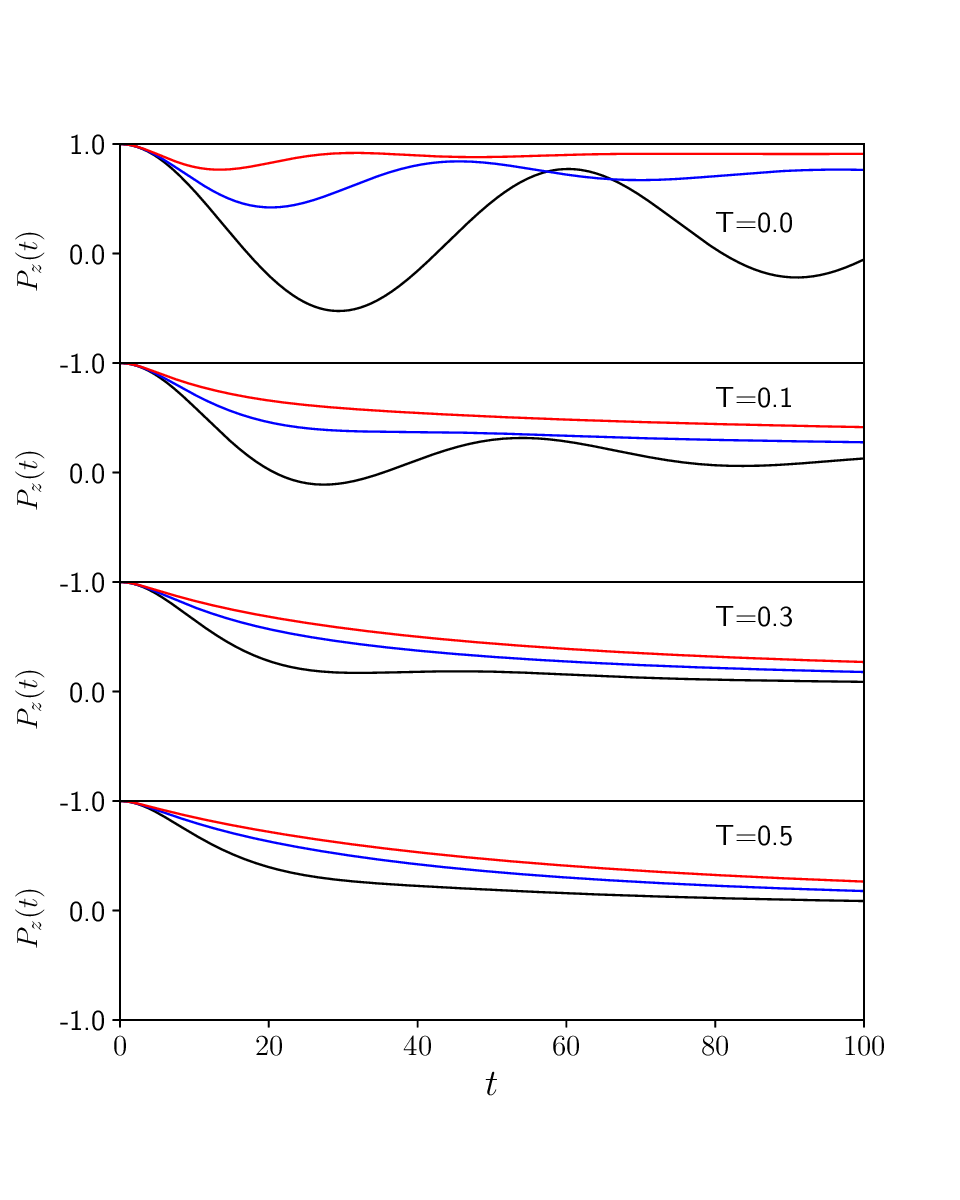}
    \caption{Spin dynamics $P_z(t)$ for the sub-Ohmic bath with $s=0.25$ and $\alpha=0.01,0.03,0.05$ (black, blue, and red lines, respectively) for several $T$ indicated in the panels.}
    \label{fig1}
\end{figure}

Fig.~\ref{fig1} displays populations $P_z(t)$  for the sub-Ohmic bath with exponent $s=0.25$ at different system-bath couplings $\alpha$ and temperatures $T$.  
At $T=0$, the oscillation frequency of $P_z(t)$ increases with increasing system-bath coupling (cf. Eq. (\ref{RT})) and the oscillatory behavior survives for larger couplings $\alpha$, in agreement with previous results. \cite{WangH2,PNalbach,KastPRL,LPSBMDiffInit} With the increase of coupling $\alpha$, $P_z(t)$ increase (that is, lie above each other) up to $t\approx 45$, in accordance with the explanations of Sec. \ref{short}. Increasing temperature acts against localization, inducing and/or fastening decay and quenching oscillations in $P_z(t)$. The bath-induced decay becomes especially pronounced in the regime of large couplings $\alpha$. Qualitatively, these dynamical behaviors of the SBM prepared under  the shifted initial condition of Eq.~(\ref{shifted}) match those of the SBM prepared under the factorized initial condition of Eq.~(\ref{normal}), but specific characteristics (e.g. periods of oscillations or decay rates) are significantly $\mu$-dependent (cf. Refs. \cite{KastPRL,KastPRB,LPSBMDiffInit}).

\begin{figure}
    \centering
    \includegraphics[width=1.0\linewidth]{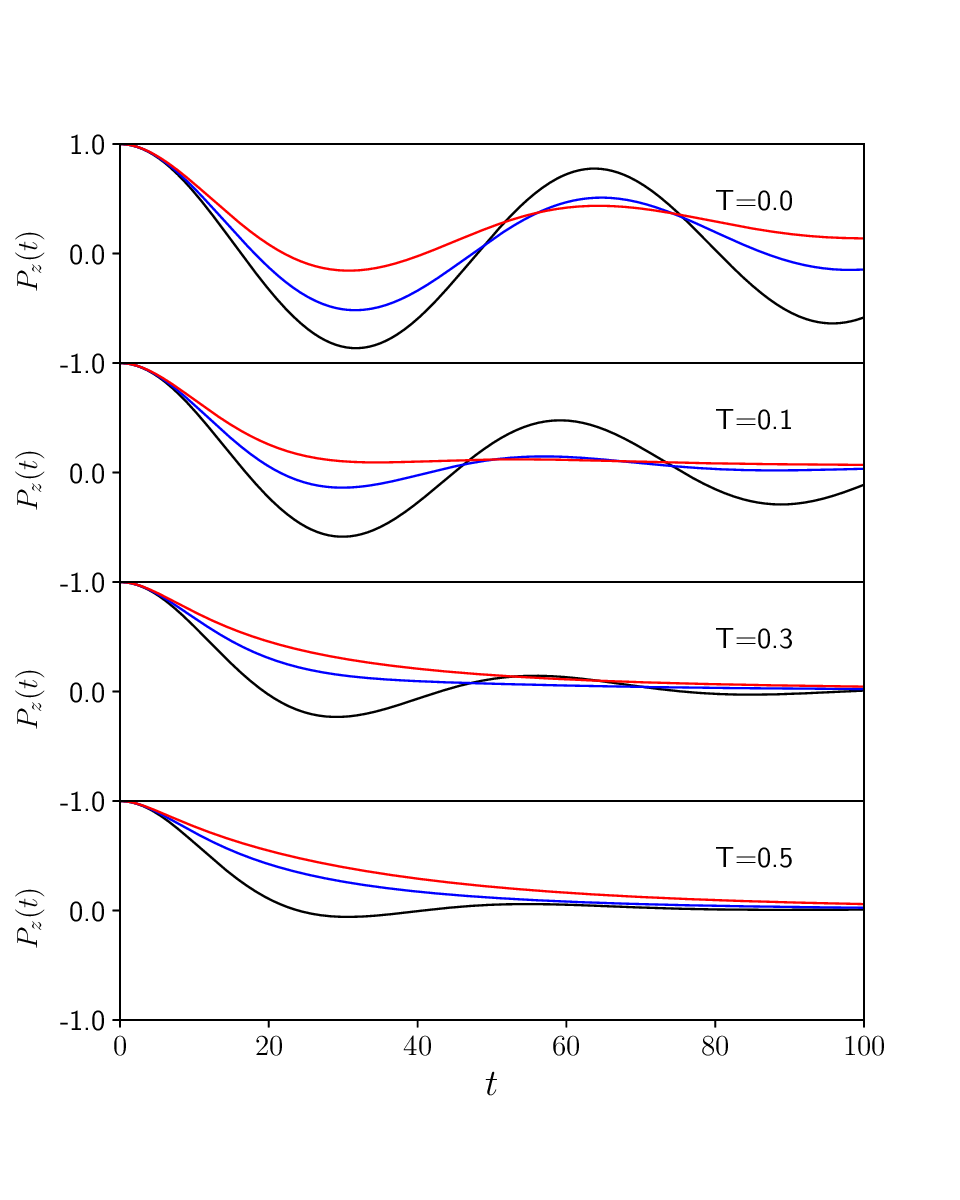}
    \caption{Same as in Fig.~\ref{fig1} but for  $s=0.5$.}
    \label{fig2}
\end{figure}

Fig.~\ref{fig2}  shows $P_z(t)$ for the sub-Ohmic bath with $s=0.5$. 
All qualitative trends for the dynamical behavior of $P_z(t)$ induced by changing the system-bath coupling $\alpha$ and temperature $T$ remain the same. However, $P_z(t)$ in Fig.~\ref{fig2} do not exhibit localization for $t\le 100$ at $T=0$ and  relatively strong coupling. In addition, the bath with $s=0.5$ is much less efficient than the bath with $s=0.25$ in quenching Rabi oscillations in the weak-coupling limit ($\alpha=0.01$).  Indeed, the black curves in Fig.~\ref{fig2} exhibit almost full recurrence at $t\approx 65$ at zero temperature, and show pronounced  underdamped oscillations at finite temperatures. The aforementioned enhancement of oscillations is in full agreement with the calculations of Ref.  \cite{LPSBMDiffInit} at $T=0$. 

\begin{figure}
    \centering
    \includegraphics[width=1.0\linewidth]{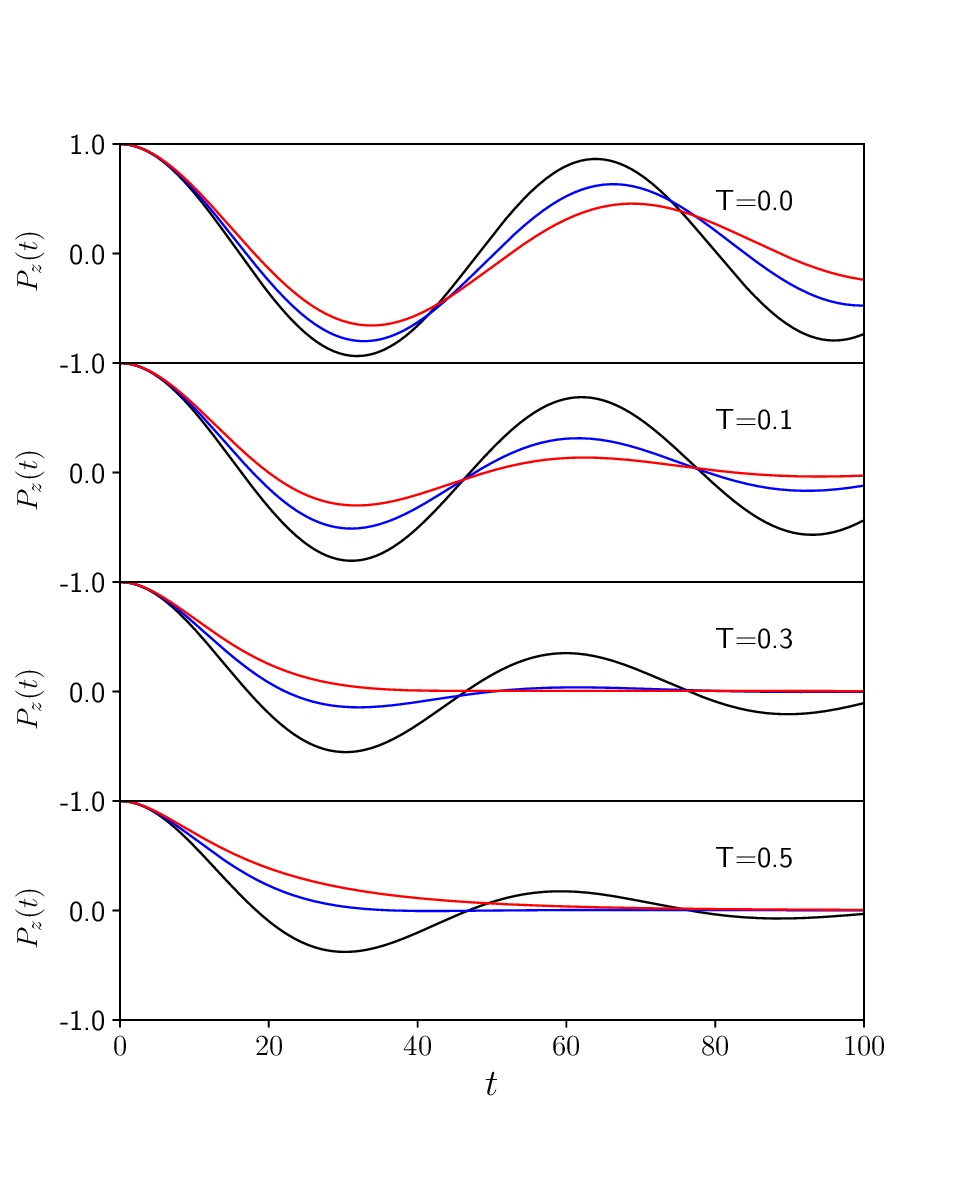}
    \caption{Same as in Fig.~\ref{fig1} but for $s=0.75$.}
    \label{fig3}
\end{figure}

Fig.~\ref{fig3} is the counterpart of Figs.~\ref{fig1} and  \ref{fig2} for $s=0.75$. The general trends in the dynamical behavior of $P_z(t)$ remain very similar. However, relaxation processes induced by the same system-bath couplings $\alpha$ are substantially $s$-dependent and, therefore,  differ in all three figures. For example, $P_z(t)$  for $\alpha=0.01$ and $T=0.5$ exhibits monotonic decay in Fig.~\ref{fig1} (incoherent regime), shows a single oscillation in Fig.~\ref{fig2} (pseudocoherent regime), and reveals  underdamped evolution in  Fig.~\ref{fig3} (coherent regime).

\begin{figure}
    \centering
    \includegraphics[width=1.0\linewidth]{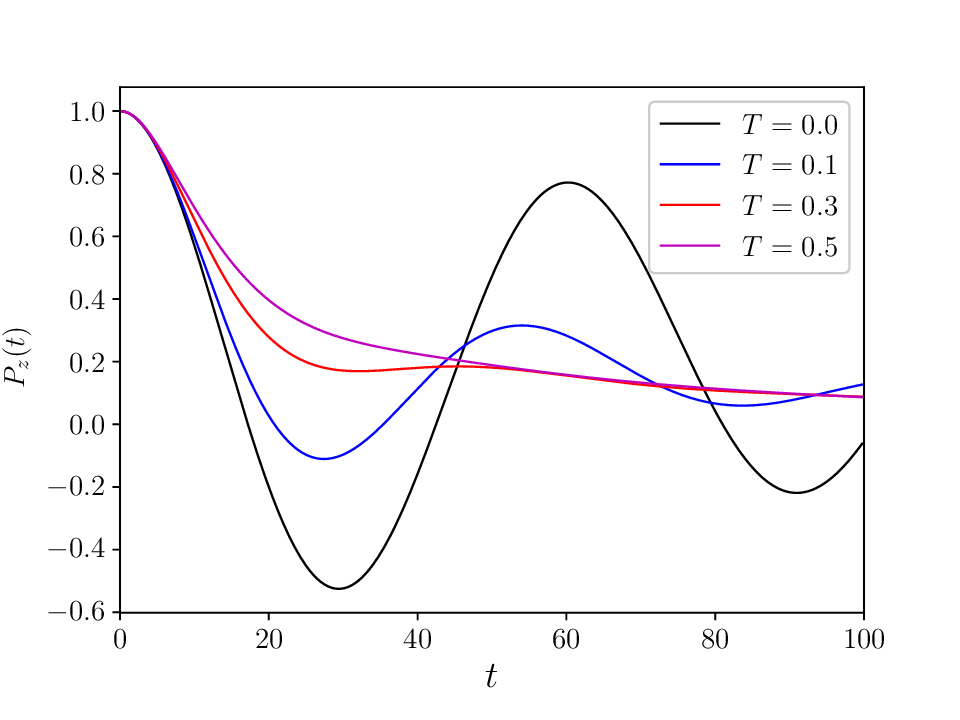}
    \caption{Spin dynamics $P_z(t)$ for the bath with  $s=0.25$ and $\alpha=0.01$ at different temperatures indicated in the legend.}
    \label{fig4}
\end{figure}

A quick perusal of Figs. \ref{fig1}-\ref{fig3} reveals that transitions between coherent, incoherent and pseudo/quasi coherent regimes are substantially temperature-dependent (see Refs. \cite{OtterpohlPRL,LPSBMDiffInit} and references therein for the precise definition of these regimes). However, a comprehensive analysis of the corresponding phase diagrams requires a painstaking characterization of $P_z(t)$ dynamics for the entire set of values of $\alpha$ and $s$, which is outside the scope of the present work. Interestingly though that increasing temperature decreases periods of Rabi oscillations in the underdamped regime. This is illustrated by Fig.~\ref{fig4}, which shows $P_z(t)$ calculated for $s=0.25$ and $\alpha=0.01$ at different temperatures. The obtained result is in full agreement with the short-time-inspired effective Rabi formulas  (\ref{RTT}) and (\ref{RT}).  

\begin{figure}
    \centering
    \includegraphics[width=1.0\linewidth]{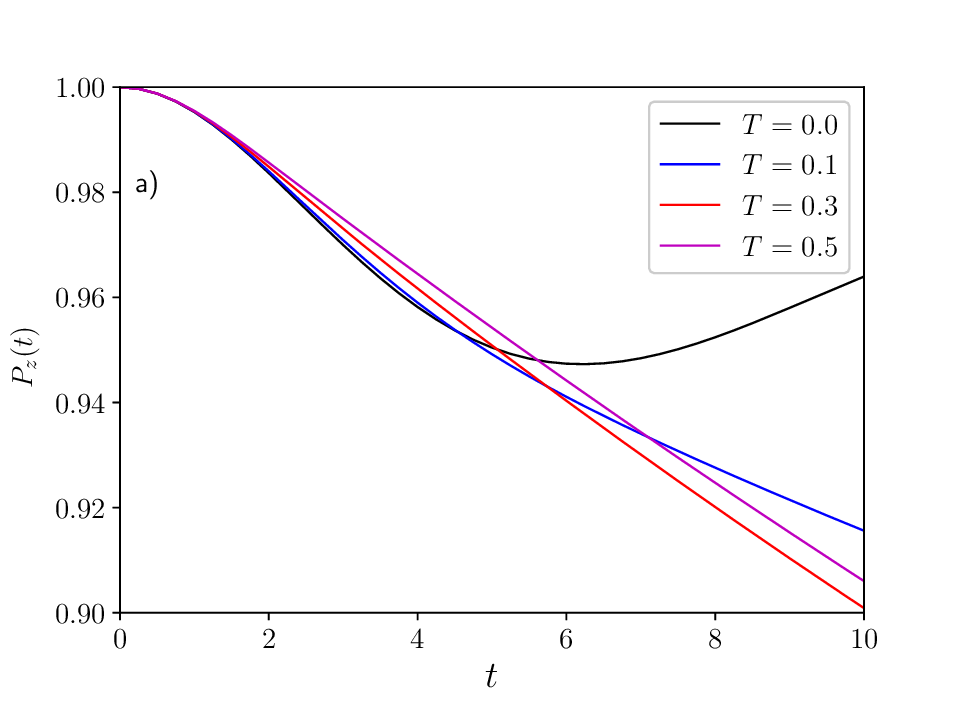}
    \includegraphics[width=1.0\linewidth]{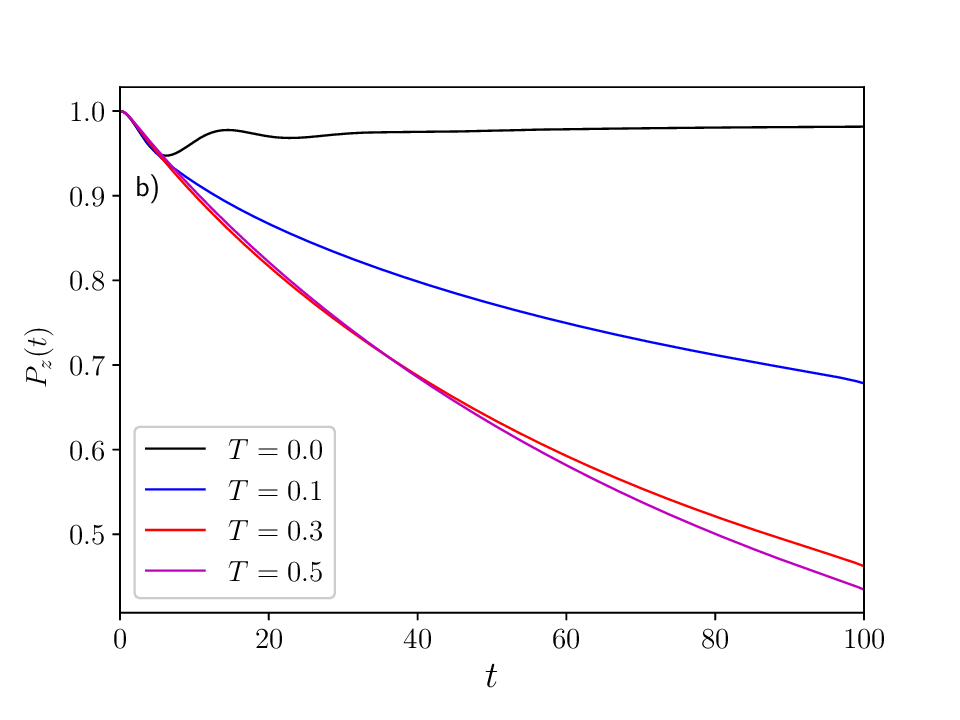}
        \caption{(a) Spin dynamics $P_z(t)$ for the bath with  $s=0.25$  and $\alpha=0.1$ at different temperatures indicated in the legends. (a) $0\le t \le 10$. (b) $0\le t \le 100$. }
    \label{fig_s025}
\end{figure}

To characterize temperature effect in more details, 
let us consider Fig.~\ref{fig_s025}. Panel (a)
shows $P_z(t)$ calculated for $s=0.25$ and $\alpha=0.1$ at different temperatures up to $t=10$. At $t<1$, $P_{z}(t)$ is temperature-independent and is solely specified by the Zeno time $\tau_{Z}=\Delta^{-1}/\sqrt{2}$, in  full agreement with Eq. (\ref{Pz}). For 
$1<t<4.5$, $P_{z}(t)$ increase with $T$ which  is also fully consistent with the prediction of Eq.~(\ref{Pz}). Then a cascade of bifurcations occurs. The black curve ($T=0$) swaps the ordering: within $4.5<t<5.5$ it changes from the lowermost to the uppermost position. The blue  curve ($T=0.1$) undergoes a similar trend, changing from the second from below to the second from above within $4.5<t<7$. The red curve ($T=0.3$) also changes the ordering, which happens at longer times. This is illustrated  by panel (b) which shows the same $P_z(t)$ but calculated up to $t=100$. Clearly, the red ($T=0.3$) and magenta ($T=0.5$) curves swap at $t\approx40$, so that the order of the curves at $t>40$ (the curves for the higher temperature  are on the bottom) is opposite to that at short times (the curves for the higher temperature are on the top).

\begin{figure}
    \centering
    \includegraphics[width=1.0\linewidth]{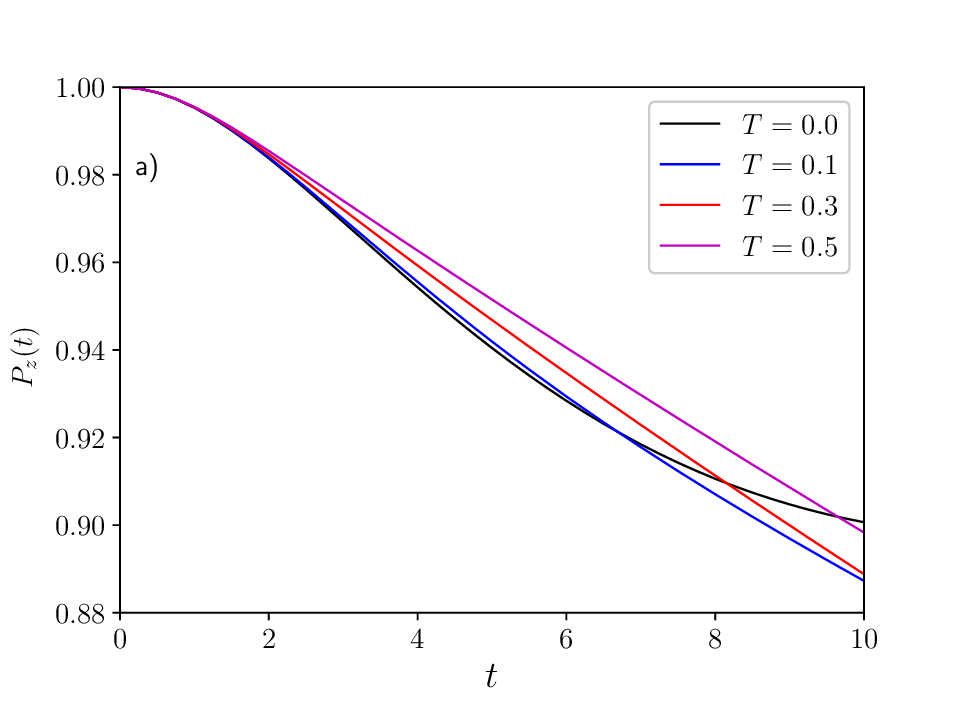}
    \includegraphics[width=1.0\linewidth]{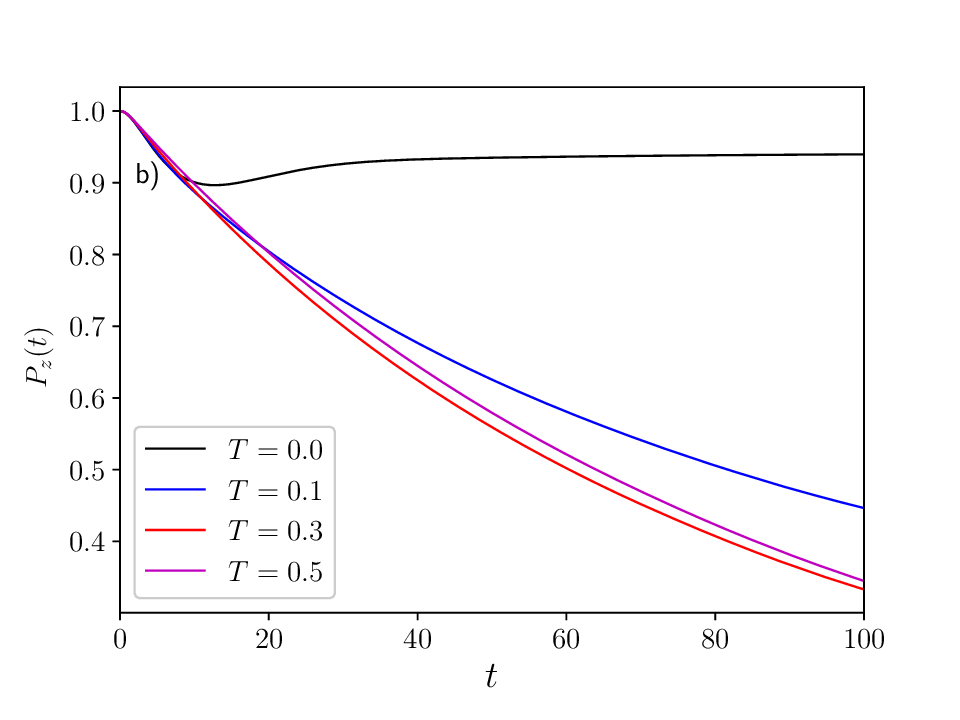}
        \caption{ Spin dynamics $P_z(t)$ for the bath with  $s=0.5$  and $\alpha=0.2$ at different temperatures indicated in the legends. (a) $0\le t \le 10$. (b) $0\le t \le 100$.}
    \label{fig_s05}
\end{figure}

The populations $P_z(t)$ evaluated for  $s=0.5$, $\alpha=0.2$ as depicted in Fig.~\ref{fig_s05} undergo similar evolutions and transformations, but at slightly different timescales. The curves in Fig.~\ref{fig_s05}(a), similar to their counterparts in Fig.~\ref{fig_s025}(a), are temperature-independent for $t<1$ and increase with temperature for $1<t<6.5$. The black curve ($T=0$)  changes from the lowermost to the uppermost position within  $6.5<t<10$. The subsequent population dynamics can be followed in Fig.~\ref{fig_s05}(b), which displays the same $P_z(t)$ on the timescale up to $t=100$. The blue curve ($T=0.1$)  changes from the second from below to the second from above within $6.5<t<20$. However, as distinct from Fig.~\ref{fig_s025}, the curves for $T=0.3$ (red) and $T=0.5$ (magenta) retain their ordering predicted by the short-time formula (\ref{Pz}).

\begin{figure}
    \centering
    \includegraphics[width=1.0\linewidth]{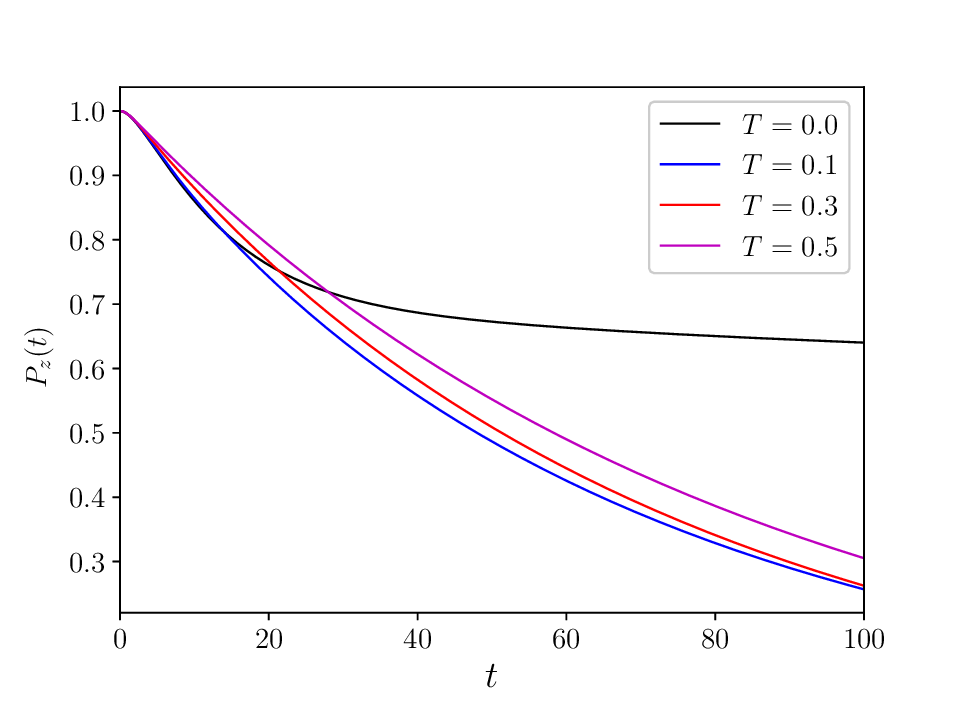}
    \caption{Spin dynamics $P_z(t)$ for the bath with  $s=0.75$  and $\alpha=0.3$ at different temperatures indicated in the legends.}
    \label{fig_s075}
\end{figure}

A different scenario is illustrated by Fig.~\ref{fig_s075} which corresponds to $s=0.75$ and $\alpha=0.3$. In this case,  the black curve ($T=0$) changes from the lowermost (as predicted by the short-time expansion of Eq.~(\ref{Pz})) to the uppermost within $0<t<28$, while the remaining curves for finite temperatures do not exhibit any change of the order, at least for the time up to $t=100$.

We thus come up with the following three-stage scenario of $P_{z}(t)$. The first, temperature-independent stage is specified by the Zeno time $\tau_Z$, which depends solely on the tunneling $\Delta$. The physical origin of this stage is well understood \cite{Zeno01,Zeno19} and is determined by the structure of the SBM (or, in a more general case, excitonic) Hamiltonian. The qualitative explanation of this behavior is  as follows. The action of the bosonic bath does not show up in the system dynamics instantaneously, and $\tau_{Z}$ is precisely the time required for the system to feel the impact of the bath.
The second stage is characterized by the increase of $P_{z}(t)$ with temperature, which can be attributed to the coupling of the bosonic DOFs to electronic DOFs (polaron effect). Indeed, Eq.~(\ref{Pz}) can be equivalently rewritten in the form of $P_{z}(t)\approx 1-2\bar{\Delta}(t,T)^{2}t^{2}$, where the effective tunneling parameter is defined as 
$\bar{\Delta}(t,T)=\Delta^{2}(1-t^{2}\Omega_{\mu}^2(T))$. Since the Rabi frequency $\Omega_{\mu}(T)$ increases with $T$, the effective tunneling $\bar{\Delta}(t,T)$ decreases with $T$, owing to the electron-vibrational (polaron) coupling. \cite{Nazir13,Lp15}  
The third stage, which may  or may not  occur for specific $\alpha$, $s$, and $T$, is specified by the cascade of bifurcations at which the populations change their short-time ordering. This third stage is therefore characterized by the decrease of $P_{z}(t)$ with    
$T$. Physically, this is a direct consequence of the increasing number of the populated bath states. Since each pair of these states contributes (with a certain weight) to the reduced system dynamics and the transition frequencies between these states are, in general, incommensurable, the summations over the increasing number of these oscillatory-in-time contributions quenchs $P_{z}(t)$. 
The bifurcation times separating the second and third stages of $P_{z}(t)$ depend significantly on the bath parameters $\alpha$ and $s$, as well as on the temperature $T$.  It is unlikely that the bifurcation phenomenon is related to the effective dynamical asymmetry discussed in Ref.~\cite{PNalbach} for the SBM  with $\epsilon=0$,  because this asymmetry is not restricted to short times but persists over the entire range of the evolution of the SBM. 

\section{Conclusions}

We employed both analytical and numerical tools to comprehensively study finite-temperature dynamics of the sub-Ohmic SBM under shifted (polarized) initial conditions. Several technical milestones have to be highlighted. First, we proved the theorem which maps the dynamics of the SBM governed by the Hamiltonian of $\hat{H}$ under shifted initial conditions onto those governed by the renormalized Hamiltonian of $\hat{H}_{\mu}$ under factorized initial conditions. Second, the theorem  facilitated the development of  efficient  and  numerically accurate method for the simulation of dynamics of the SBM under shifted initial conditions. The method is based on the powerful HEOM-TSTT  integrator designed for the simulation of multidimensional quantum systems under the factorized initial condition. Third, the theorem  was used for the evaluation of first few terms in the Taylor-series
expansion of the population $P_{z}(t)$, which substantially aided the interpretation of the numerics.

The main results of our simulations can be briefly summarized as follows. First, we demonstrated high sensitivity of the dynamics to both initial preparation of the bath and temperature. Second, we gave the benchmark results which could be used for testing other, accurate and approximate, simulation methods and protocols. Third, we discovered  the bifurcation phenomenon, which separates two regimes of the dynamics in the time domain. Before the bifurcation time, elevated temperatures slow down $P_{z}(t)$. After the bifurcation time, they accelerate $P_{z}(t)$  .

We predicted that the established bifurcation phenomenon is general and can be found in other dissipative systems. Furthermore, it  
is potentially of great interest, because it demonstrates that the initial preparation of the bath (polarized bath + temperature) allows us to efficiently manipulate depopulation rates. Taking into account recent progress in engineering and/or emulation of bosonic baths with arbitrary spectral densities, \cite{Home15,Ustinov23,Pollanen23}  the bifurcation phenomena may open up exciting perspectives for the optimization and the control of nanosystems.

\begin{acknowledgements}
L. P. Chen  acknowledges support from the Key Research Project of Zhejiang Lab (No. 2021PE0AC02). M. F. G. acknowledges support
from the National Natural Science Foundation of China
(Grant No. 22373028).
H. T. and R. B. acknowledge the support by the Spoke 7 "Materials and Molecular Sciences" of ICSC – Centro Nazionale di Ricerca in High-Performance Computing, Big Data and Quantum Computing, funded by European Union – NextGenerationEU.
\end{acknowledgements}

\section*{Data availability}
Further data that support the findings of this study are available from the corresponding author upon reasonable request.

%\bibliography{RBlibrary,esprit}

%%%%%%%%%%%%%%%%%%%%%%%%%%%%%%%%%%%%%%%%%%%%%%%%%%%%%%%%%%%%%%%%%%%%%

\end{document}